# Appearance of Antiferromagnetism and Superconductivity in Superconducting Cuprates

by

Krzysztof Kucab, Grzegorz Górski and Jerzy Mizia

Institute of Physics, University of Rzeszów
Al. Rejtana 16A, 35-959 Rzeszów, Poland

**Abstract**

We represent the superconducting ceramic compounds by the single band extended Hubbard model. We solve this model for the simultaneous presence of antiferromagnetism and the $d$-wave superconductivity in the Hartree-Fock (H-F) and in the coherent potential (CP) approximation, which is applied to the on-site Coulomb repulsion $U$.

The hopping interaction used in addition to the Coulomb repulsion causes rapid expansion of the band at carrier concentrations departing from unity. This expansion shifts the $d$-wave superconductivity away from the half filled point reducing its occupation range approximately to the experimental range in the YBaCuO compound.

The CP approximation used for the Coulomb repulsion is justified by the large ratio of Coulomb repulsion to the effective width of perturbed density of states being reduced by the hopping interaction.

Fast disappearance of antiferromagnetism observed experimentally is supported by the relatively large value of the total width of unperturbed density of states.

It is shown that our model is capable of describing the electron doped compounds as well.

PACS codes: 71.10.Fd, 74.20.-z, 75.10.Lp, 74.25.Dw, 75.50.Ee

Keywords: equation of motion approach, Hubbard model, antiferromagnetism, superconductivity

# I. Introduction

The coexistence of superconductivity (SC) and antiferromagnetism (AF) phenomena was observed in many types of superconducting layered cuprates like $Nd_{2-x}Ce_xCuO_4$, $La_{2-x}Sr_xCuO_4$ [1,2] or $YBa_2Cu_3O_{6+x}$, in organic superconductors [3], in Uranium- and Cerium-based heavy-fermion systems [4], in Fe-based oxygen pnictide [5] and $MFe_2As_2$ family [6,7]. Transition between these two phases depends on carrier doping (by holes or electrons) and on external pressure. In layered cuprates, Fe-based oxygen pnictides, and iron arsenide family the superconducting and antiferromagnetic state can be suppressed by chemical doping. Almost all high-temperature superconducting cuprates at half-filling are the antiferromagnetic insulator. Hole doping (in $La_{2-x}Sr_xCuO_4$ and $YBa_2Cu_3O_{6+x}$) or electron doping ($Nd_{2-x}Ce_xCuO_4$) causes vanishing of the antiferromagnetic state and appearance of superconducting state in the metallic state. The main difference between the hole-doped and electron-doped cuprates is in the range of existing antiferromagnetic state. In the hole-doped cuprates AF vanishes already at small doping (~0.03) and before appearance of SC state. In the electron-doped cuprates the AF state exists up to the relatively high values of doping (~0.15), and the SC state rises before the AF state is extinguish. Schematic diagram for the hole-doped and electron-doped superconducting cuprates is shown in Fig. 1. Phase diagram of $YBa_2Cu_3O_{6+x}$ (as a function of carrier concentration) compound is very similar to the diagram of $La_{2-x}Sr_xCuO_4$ compound after interchanging the dopant concentration to the hole number by the relation: $x = 1 - n$ (see e.g. [8]), and is shown in Fig. 6 below.

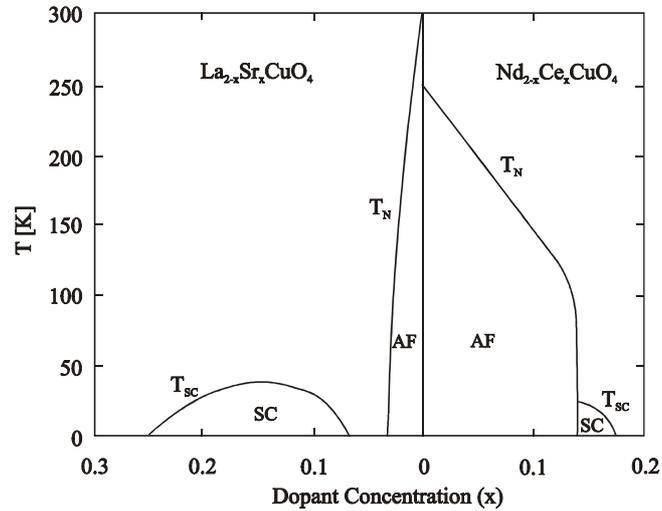

Fig. 1 Schematic phase diagram for both electron-doped ($Nd_{2-x}Ce_xCuO_{4-y}$) and hole-doped ($La_{2-x}Sr_xCuO_{4-y}$) superconductors, showing antiferromagnetic (AF) and superconducting (SC) phases.

There are also numerous theoretical papers devoted to the problem of competition between AF and SC. Several methods are used to describe this phenomenon. The first attempts were done by use of the mean-field approximation ([9-11]). Some authors carry out their computations using the dynamical mean-field theory (DMFT), which is based on the transition from the Hubbard model to the simple impurity Anderson model (SIAM) [12].

An early version of cluster DMFT was used by Lichtenstein [13]. The authors found that *d*-wave superconductivity and antiferromagnetism coexist over most of the doping range. In the paper [14] the cellular dynamical mean-field theory (CDMFT) was used to compute zero-temperature properties of the two-dimensional square lattice Hubbard model (using the exact-diagonalization method). The authors have calculated the phase diagram that describes the competition between antiferromagnetism and superconductivity for both hole and electron doping at various values of $U$. They found, in general, homogeneous coexistence between AF and *d*-wave SC in the underdoped region (that feature is quite generally observed in quantum cluster methods, variational approaches, and mean-field theories that do not allow for spin or charge density modulations on large length scales). There is also another fully causal self-consistent method i.e. the dynamical cluster approximation (DCA) (see e.g. [15]). Using this method one can reduce the complexity of the lattice problem by mapping it to a finite-size cluster self-consistently embedded in a mean field. The main difference with their classical counterparts arises from the presence of quantum fluctuations. Mean-field theories for quantum systems with itinerant degrees of freedom cut off spatial fluctuations but take full account of temporal fluctuations. As a result the mean field is a time- or frequency dependent quantity.

Another methods use the spectral density approach [16]. The results obtained by the help of this method show the growing range of AF state with increasing $U$ (the opposite behavior compared to the methods described earlier).

Despite the large progress in theoretical analysis of coexistence between SC and AF there is still the need for description of experimental results. Particular attention should be devoted to decreasing the concentration range of coexistence between those two phases. Important is also explaining the strong electron-hole asymmetry especially visible for the AF ordering (see Fig. 1).

The main objective of this paper is to explain simultaneous appearance of AF and the *d*-type SC as shown by the diagram in Fig. 1, using the itinerant extended Hubbard model described by Hamiltonian of Eq. (1) below.

**II. Model Hamiltonian**

Our single band Hamiltonian has to represent the complex situation in the $CuO_2$ plane of layered compounds. Therefore it includes the on-site Coulomb repulsion $U$ and the inter-site charge-charge interaction $V$. In addition we have the hopping interaction $\Delta t$ (see [17-19]). This interaction causes the rapid growth of the bandwidth when the electron occupation *n* departs from unity. It stimulates the *s* type superconductivity (see [17,18]), which in our model is suppressed by the relatively strong effective Coulomb repulsion. The existing *d*-wave superconductivity (see [20-22]) is created by the negative inter-site charge-charge interaction $V$, and it is not hindered by the repulsion $U$. The concentration dependence of the effective bandwidth, $D(n)$, due to the nonzero value of $\Delta t$ interaction, enhances the $U/D(n)$ ratio at growing occupation. This in turn in the CP approximation causes the repulsion of the *d*-wave SC state away from the half filled point. The model Hamiltonian used in our work is presented by Eq. (1).

$$H = -\sum_{<ij>\sigma}\left[t-\Delta t\left(\hat{n}_{i-\sigma}+\hat{n}_{j-\sigma}\right)\right]c^+_{i\sigma}c_{j\sigma} - \mu_0\sum_{i\sigma}\hat{n}_{i\sigma} + \frac{U}{2}\sum_{i\sigma}\hat{n}_{i\sigma}\hat{n}_{i-\sigma} + \frac{V}{2}\sum_{<ij>}\hat{n}_i\hat{n}_j - F_{in}\sum_{i\sigma}n_\sigma\hat{n}_{i\sigma}. \quad (1)$$

In the above equation $\mu_0$ is the chemical potential, $\hat{n}_{i\sigma} = c^+_{i\sigma}c_{i\sigma}$ is the electron number operator, $c^+_{i\sigma}$ ($c_{i\sigma}$) creates (annihilates) an electron with spin $\sigma$ on the *i*-th lattice site, and $n_\sigma$ is the probability of finding the electron with spin $\sigma$ in a band (we assume for simplicity, that the band is fully degenerated, i.e. the single band is composed of identical orbitals). Parameter *t* is the hopping amplitude for an electron of spin $\sigma$ when both sites *i* and *j* are empty, and $\Delta t$ is the hopping interaction. This interaction is defined as $\Delta t = t - t_1$, where $t_1$ is the hopping amplitude for an electron of spin $\sigma$ when one of the sites *i* or *j* is occupied by an electron with opposite spin [20]. The single site Hund's type exchange interaction $F_{in}$ added in the Hamiltonian can be interpreted as the interaction between different orbitals on the same lattice site in the multi-orbital single band model.

By the reduction of the three-band extended Hubbard model to the effective single-band model used by us, we obtain the negative value of inter-site charge-charge interaction $V$. In the three-band extended Hubbard model the effective inter-site charge-charge interaction can take the negative value as according to Weber [22] at relatively strong $U/D_0$ ($D_0$ is the half-bandwidth) it can be expressed by the formula

$$V = V_z - V_x < 0 , \quad (2)$$

where $V_z$ and $V_x$ are the interactions between the copper $3d_{3z^2-r^2}$ and $3d_{x^2-y^2}$ orbitals and the oxygen orbital $2p$ in the CuO$_2$ plane, respectively.

The analysis of coexistence between AF and SC is carried using equations of motion for the Green functions.

In the further analysis of the Hamiltonian (1) we divide the crystal lattice into two interpenetrating magnetic sub-lattices $\alpha, \beta$. The two interpenetrating sub-lattices generate the $\alpha\beta\alpha\beta$ (G-type) structure of antiferromagnetism, in which the atoms on sub-lattice $\alpha$ have as their nearest neighbors only the atoms on sub-lattice $\beta$. The operator $\hat{n}_{i\sigma} = c^+_{i\sigma}c_{i\sigma}$ is becoming $\hat{n}^\gamma_{i\sigma} = \gamma^+_{i\sigma}\gamma_{i\sigma}$ for sites $\gamma = \alpha, \beta$. We apply to the inter-site interactions $\Delta t$ and $V$ the full H-F approximation, which includes the averages of the inter-site type $I_{AF} \equiv \langle c^+_{i\sigma} c_{j\sigma} \rangle$. We consider possible superconducting $d$-wave and antiferromagnetic ordering at the same time.

After using the above assumptions and approximations we can write the Hamiltonian in the following form

$$H = H_0 + H_{SC} + H_I, \qquad (3)$$

where the unperturbed part of the Hamiltonian is given by

$$H_0 = -t_{eff} \sum_{\substack{<ij> \\ \sigma}} \left(\alpha^+_{i\sigma}\beta_{j\sigma} + \beta^+_{j\sigma}\alpha_{i\sigma}\right) - \mu \sum_{\substack{i,\gamma,\sigma \\ (\gamma=\alpha,\beta)}} \hat{n}^\gamma_{i\sigma} + \sum_{\substack{i,\gamma,\sigma \\ (\gamma=\alpha,\beta)}} \Sigma^\sigma_\gamma \hat{n}^\gamma_{i\sigma}, \qquad (4)$$

the perturbed part is equal to

$$H_I = \sum_{\substack{i\sigma \\ (\gamma=\alpha,\beta)}} \left(U\hat{n}^\gamma_{i-\sigma} - F_{in}n^\sigma_\gamma - \Sigma^\sigma_\gamma\right)\hat{n}^\gamma_{i\sigma}, \qquad (5)$$

and the superconducting part has the form

$$H_{SC} = \sum_{i,\sigma}\left[\left(2z\Delta t\Delta_{\alpha\beta} + U\Delta_0\right)\gamma^+_{i\sigma}\gamma^+_{i-\sigma} + h.c.\right] + \sum_{\substack{<ij> \\ \sigma}}\left[\left(\Delta t\Delta_0 + \frac{1}{2}V\Delta_{\alpha\beta}\right)\sigma\alpha^+_{i\sigma}\beta^+_{j-\sigma} + h.c.\right]. \quad (6)$$

The effective hopping interaction appearing in Eq. (4) is given by

$$t_{eff} = tb^{AF}, \quad \text{where} \quad b^{AF} = 1 - \frac{\Delta t}{t}n + \frac{V}{t}I_{AF}, \qquad (7)$$

and the effective chemical potential is equal to

$$\mu = \mu_0 - zVn - 2z\Delta tI_{AF}. \qquad (8)$$

The operators $\alpha^+_{i\sigma}(\alpha_{i\sigma})$, $\beta^+_{i\sigma}(\beta_{i\sigma})$ create (annihilate) electrons with spin $\sigma$ on sub-lattices $\alpha, \beta$, $z$ is the number of nearest neighbors, $n^\sigma_\gamma$ is the average electron occupation on sites $\gamma = \alpha, \beta$, and $\Sigma^\sigma_\gamma$ is the self energy for electrons with spin $\sigma$ on sites $\gamma$.

The carrier concentration and antiferromagnetic moment are given as

$$n = n^\sigma_\alpha + n^{-\sigma}_\alpha = n^{-\sigma}_\beta + n^\sigma_\beta, \qquad (9)$$

$$m = n^\sigma_\alpha - n^{-\sigma}_\alpha = n^{-\sigma}_\beta - n^\sigma_\beta, \qquad (10)$$

and the superconducting order parameters have the form

$$\Delta_{\alpha\beta} = \frac{1}{2}\sum_{\sigma}\sigma\langle\beta_{j-\sigma}\alpha_{i\sigma}\rangle, \quad \Delta_0 = \langle c_{i\downarrow}c_{i\uparrow}\rangle. \tag{11}$$

The self-energy $\Sigma_\gamma^\sigma$ which appears in Eqs. (4) and (5) is calculated by treating expression (5) as a perturbation.

Subsequently we will use the equation of motion for the Green functions to analyze the Hamiltonian (3). In the Appendix A we derive the formalism for the interaction $U$ (see Eq. (5)) treated in the H-F approximation. The final relations for carrier concentration, AF magnetization and the energy gap of the *d*-wave SC state are given by Eqs. (A.11), (A.12) and (A.19) in Appendix A.

To obtain results closer to the experiment we will apply later on the CP approximation to the interaction $U$ in Eq. (5). Outline of the derivations is presented in the Appendix B. The results (analogous to these of H-F approximation) are given by Eqs. (B.5), (B.6) and (B.11) of Appendix B.

### III. Numerical Results

**(A) The Hartree-Fock calculations**

In this Section we apply the H-F approximation to the on-site Coulomb repulsion $U$. In addition we introduce the inter-site charge-charge interaction $V$ (creating the SC of the *d* type) to which we also apply the H-F approximation. In the H-F approximation the antiferromagnetism is created by the sum of the intra-atomic exchange field, $F_{in}$, and the on-site Coulomb repulsion $U$. In our single band model the intra-atomic exchange field can be originated by the inter-orbital exchange interaction on the same site of the multi orbital model. The influence of the hopping interaction $\Delta t$ for critical interaction creating the AF state in the H-F approximation, based on the 2D tight binding initial density of states (DOS), is shown in Fig. 2. This interaction causes the bandwidth change with occupation. When we depart from the half-filled point the exchange field is becoming less capable of creating AF since the critical field necessary to sustain AF grows. We may compare these results with the CP approach in Fig. 9.

The results shown in Fig. 2 indicate that the calculated range of AF state is larger than the experimental range in the doped YBaCuO compound; moreover the hopping interaction causes increase of this theoretical range.

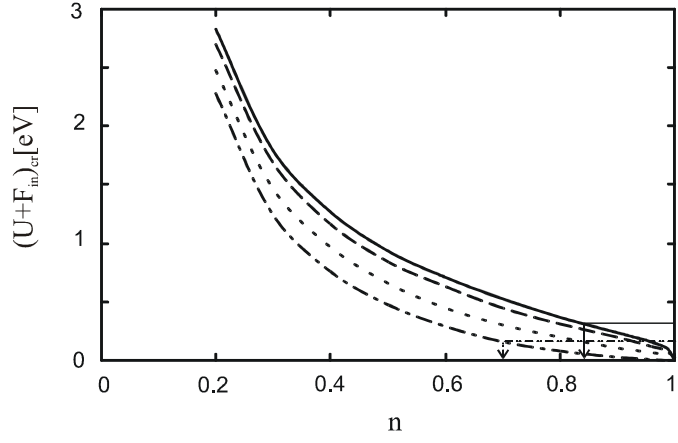

Fig.2 Critical curves for the antiferromagnetic interaction $(U + F_{in})_{cr}$ in the Hartree-Fock approximation. Calculations are performed for the two dimensional tight binding DOS. The parameters are: $V = -0.09$eV, and $D_0 = 0.5$eV. The hopping interactions are: $\Delta t = 0$ – solid line, $\Delta t = 0.2t$ – dashed line, $\Delta t = 0.6t$ – dotted line, $\Delta t = 0.99t$ – dot-dashed line. The $U + F_{in}$ values fitted to the Néel's temperature of 500K at $n = 1$ are shown by the horizontal lines for $\Delta t = 0$ and $\Delta t = 0.99t$. The intersections with critical curves determine the carrier occupation range of existing AF state in these two cases.

In Fig. 3 below we show the phase diagram obtained by the help of equations presented in Appendix A. The Néel's temperature was calculated from Eq. (A.12) in the limit of $m \to 0$, and the SC critical temperature from Eq. (A.19) in the limit of $\Delta_k \to 0$. Both the SC and AF critical temperatures were obtained using in addition the Eq. (A.11), which allows for obtaining the chemical potential.

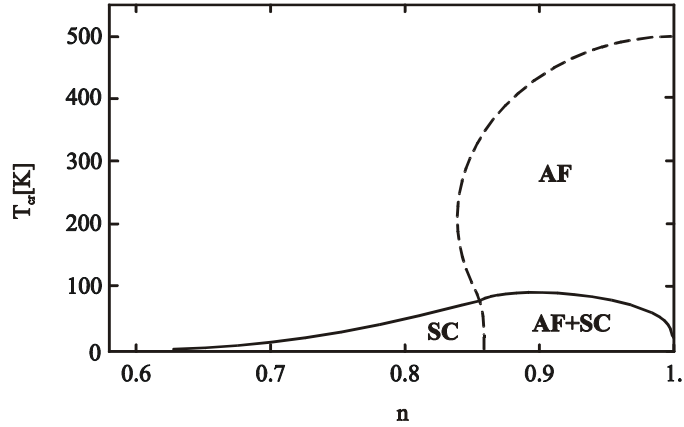

Fig. 3 The dependence of superconducting critical temperature $T_{SC}$ (solid line) and Néel's temperature $T_N$ (dashed line) on carrier concentration $n$ for the SC energy gap of the $d$-type. The H-F approximation was applied to interactions $U$ and $V$. The interactions are: $V = -0.09$eV, $U + F_{in} = 0.32$eV, the half-bandwidth $D_0 = 0.5$eV and $n$ is the carrier concentration. The hopping interaction $\Delta t = 0$.

As one can see from Fig. 3, in the H-F approximation there is a broad range of carrier concentration at which the AF state coexists with SC state, without repelling each other. The SC state disappears merely at concentration $n \to 1$. This is the consequence of using the low level approximation for the interactions. As can be seen from Fig. 2 the range of occurrence of the AF state is determined by the value of $U + F_{in}$ which gives the Néel's temperature at $n = 1$. The AF state disappears for the concentrations for which the value of $U + F_{in}$ becomes equal to $(U + F_{in})_{cr}$.

**(B) Coherent potential calculations, the bandwidth independent of carrier concentration ($\Delta t = 0$)**

The phase diagram obtained in the H-F approximation shows the broad range of coexistence between SC and AF phases. This result disagree with the experimental evidence for cuprates, where the range of coexistence is either very narrow (e.g. for $Nd_{2-x}Ce_xCuO_4$) or it does not exist (e.g. for $YBa_2Cu_3O_{6+x}$). Wysokiński and Domański [21] have shown that in the higher order than the H-F approximations the maximum of the critical SC temperature is shifted to the concentrations below the half-filling. At large enough value of $U$ they obtained disappearance of SC phase at $n = 1$. Therefore, we will apply the coherent potential (CP) approximation to the Coulomb interaction (see e.g. [20,23]) to analyze coexistence of SC and AF phases (see Appendix B). The inter-site charge-charge interaction $V$ we still treat in the H-F approximation. It is important to note here, that in this stage our computations are still made with $\Delta t = 0$, i.e. with the bandwidth independent on carrier concentration. The interaction $\Delta t \neq 0$ will be considered in the next step.

First we will investigate the influence of the bandwidth and Coulomb repulsion $U$ on the range of AF state. The numerical results show, that the increase of the initial bandwidth causes dramatic decrease in the range of AF state (see Fig. 4). For further calculations we adopt the value of half-bandwidth $D_0 = 2\text{eV}$. Such a bandwidth can be justified by the values of hopping integral obtained by Feiner et al. [24] in the effective single-band model, which came about by reducing the three-band extended Hubbard model representing $CuO_2$ planes in the high-$T_c$ cuprates.

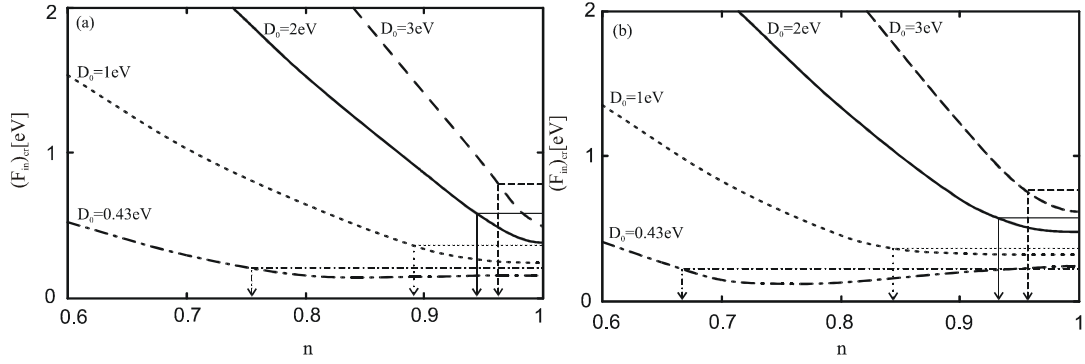

Fig. 4 The critical curves for the exchange on-site antiferromagnetic interaction $\left(F_{in}\right)_{cr}$ in function of carrier concentration $n$ for different values of half-bandwidth $D_0$. The Coulomb on-site repulsion $U$ is treated in the CP approximation. The $F_{in}$ values fitted to the Néel's temperature at $n=1$ are shown in this figure by the horizontal lines. At the intersection with critical curves they determine the range of AF state. The parameters for both figures are: $V = -0.21\text{eV}$, and $\Delta t = 0$. Fig. (a) is for $U = 0.17\text{eV}$ and Fig. (b) is for $U = 0.3\text{eV}$.

As we can see, the influence of the Coulomb repulsion $U$ on the range of the AF state is relatively small comparing to the influence of the initial unperturbed bandwidth. Therefore the mechanism controlling the range of AF state is the increase in $D_0$, and (as we will see in the next Section) we can control the range of SC state by using the proper $U$ values (which do not strongly influence the AF state).

In Fig. 5 below we show the dependence of Néel's temperature on electron concentration for different $U$ values at $D_0 = 2\text{eV}$ (interaction $\Delta t = 0$). One can see the growing range of AF state with increasing $U$. This effect is observed also in the spectral density approach [16]. Opposite effect was obtained by Kancharla et. al [14] who introduced the super exchange coupling $J = 4t^2/U$, in which case the increase of $U$ decreased a value of $J$ leading to the decrease of AF range.

The range of AF state obtained by us is relatively high; it goes beyond the experimental evidence for $La_{2-x}Sr_xCuO_4$ or $YBa_2Cu_3O_{6+x}$. It is due to the fact that in those compounds the AF phase exists in the insulating phase, while at our values of $U$ we have metallic phase. Broader range of AF ordering in the metallic phase exists in the electron-doped $Nd_{2-x}Ce_xCuO_4$ and in the five-layered $HgBa_2Ca_4Cu_5O_{12-x}$ [25]. To obtain a Mott insulator at $n = 1$, we will employ in the next chapter the interaction $\Delta t \neq 0$ which will cause the band-split, due to the decrease of effective $D$ and increase of $U/D$, specially for carrier concentrations close to unity.

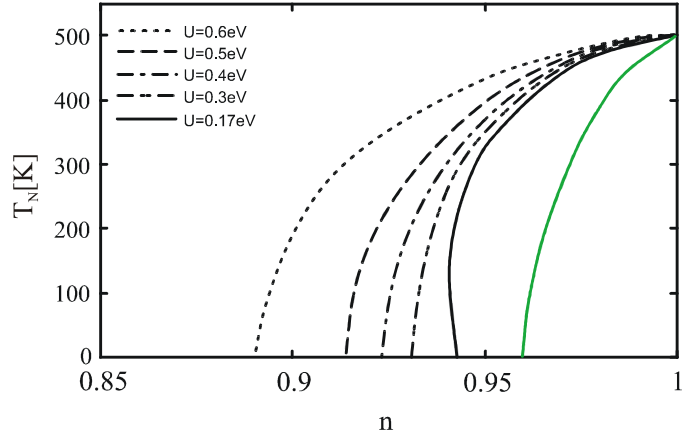

Fig. 5 The dependence of critical AF temperature on electron concentration for different $U$ values at $D_0 = 2\text{eV}$, (interactions $\Delta t = 0$, $V = -0.21\text{eV}$). Experimental Néel's temperature at $n = 1$ was obtained by fitting the $F_{in}$ parameter. The green curve represents the experimental data for YBCO.

In Fig. 6 we show phase diagram for the antiferromagnetic and superconducting states. At first we assumed the parameters $D_0 = 0.43\text{eV}$ and $U = 0.17\text{eV}$. Values of the charge-charge interaction and on-site exchange interaction were fitted to the maximum of 100 K for superconductivity and the Néel's temperature of the order 500 K, respectively. As these interactions preserve the electron-hole symmetry we show the results only for the hole-doped concentration range ($n \leq 1$). The results for the electron-doped domain ($n \geq 1$) are identical. They show strong competition between AF ordering and the $d$-wave SC. The influence of AF on SC is evident by observing that the SC temperature is rapidly dropping to zero after meeting with the AF state boundary line. There is also the opposite effect of limiting the AF range by the $d$-wave SC. As a result calculated range of coexistence between AF and SC state is very narrow. This is very different than the result of a H-F calculations presented in the previous subsection, where the coexistence took place up to $n = 1$.

At $D_0 = 0.43\text{eV}$ and $U = 0.17\text{eV}$ both the range of SC and of AF are very different from the experimental evidence. Analyzing the results presented in Figs. 4a and 5 we can see that at parameters $D_0 = 2\text{eV}$ and $U = 0.17\text{eV}$ the range of existing AF state is close to the experimental data for the $YBa_2Cu_3O_7$ compound, therefore in Fig. 7 we show results calculated with these parameters.

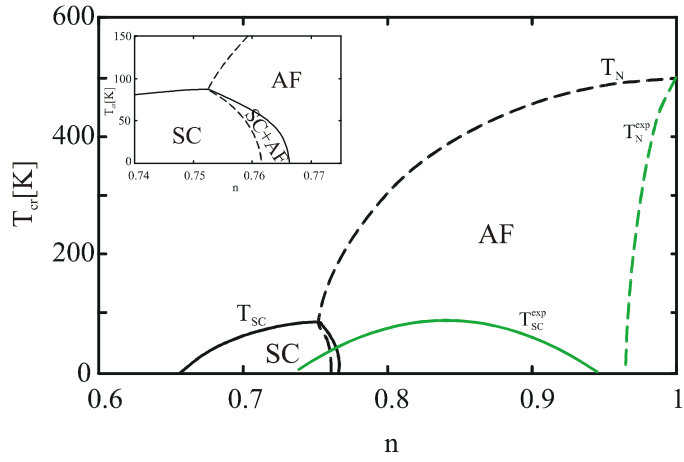

Fig. 6 The dependence of the calculated superconducting critical $d$-type temperature $T_{SC}$ (black solid line) and Néel's temperature $T_N$ (black dashed line) on carrier concentration $n$. CP approximation was applied to $U$. The interactions are: $V = -0.18 \text{eV}$, $F_{in} = 0.21 \text{eV}$, $\Delta t = 0$ and $U = 0.17 \text{eV}$; the value of half-bandwidth: $D_0 = 0.43 \text{eV}$. The inset shows the region of coexistence between AF and SC. Experimental data are shown by the green curves.

In Fig. 7 the theoretical results for the larger $D_0 = 2 \text{eV}$ are shown together with the experimental data for the YBa$_2$Cu$_3$O$_7$ compound. As one can see the calculated $d$-wave superconductivity would remain up to the half filled point but it is suppressed by the antiferromagnetism. The antiferromagnetism is closer to the experimental narrow range for YBCO compound than in the case of $D_0 = 0.43 \text{eV}$, in which the theoretical AF ordering persisted to concentrations much lower than the experimental AF state. Now at $D_0 = 2 \text{eV}$ the AF state is limited to more realistic range.

The phase diagram seen in Fig. 7 has a character closer to the electron doped ceramics based on Neodymium, where the SC state is entering the AF state at its maximum critical temperature. This would suggest the small role of $\Delta t$ interaction in these ceramics as the diagram of Fig. 7 was constructed at $\Delta t = 0$.

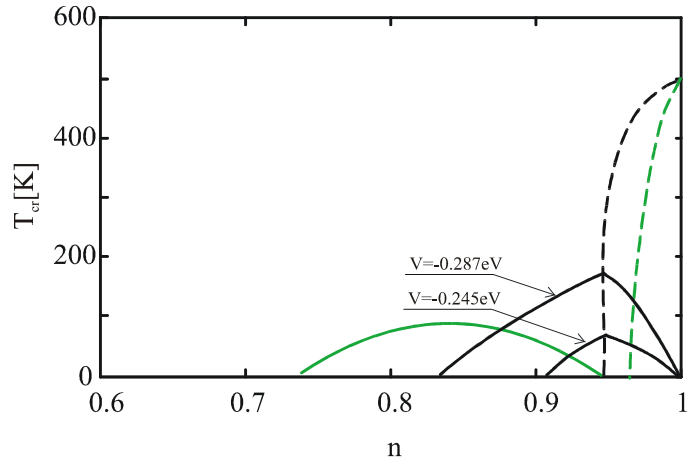

Fig. 7 Dependence of the calculated superconducting critical temperature $T_{SC}$ for the SC energy gap of the *d*-type (black solid line) and Néel's temperature $T_N$ (black dashed line) on carrier concentration $n$ (CP approximation was applied to $U$). The interactions are: $F_{in} = 0.6\text{eV}$, $\Delta t = 0$ and $U = 0.17\text{eV}$; the value of half-bandwidth is $D_0 = 2\text{eV}$. Experimental data for YBCO are shown by the green curves.

**(C) Coherent potential calculations, the carrier concentration dependent bandwidth ($\Delta t \neq 0$)**

To rectify the situation of the *d*-wave SC persisting almost to the half filled band we include now the hopping interaction $\Delta t$. This interaction was broadly used to the description of superconducting state [17,20,26,27], metal-insulator transition [28] and in magnetism [29-31]. Its influence on superconductivity goes in two ways. It creates the pairing potential for the *s*-wave SC at concentrations near $n = 0$ and $n = 2$ [17,27], and it changes the bandwidth what shifts the SC phase of the *d*-wave type into the more reliably range of concentrations (see Eq. (7) for the change of the bandwidth with interaction $\Delta t$ and $V$). The influence of the hopping interaction on the bandwidth change can be observed in the diagrams of Fig. 8 presenting the paramagnetic density of states in function of energy for different carrier concentrations. They are calculated from the initial (without interactions) two dimensional tight binding DOS. As we can see, the hopping interaction $\Delta t$ causes decrease of the sub-band widths. This decrease increases with carrier concentration $n$ and at $n \approx 0.93$ the band becomes split into two sub bands. At half filling we have the insulating phase (band split) and the lower Hubbard band is becoming very narrow due to the very small effective hopping integral in the presence of another electron $t_1 \approx 0$ (see [32]).

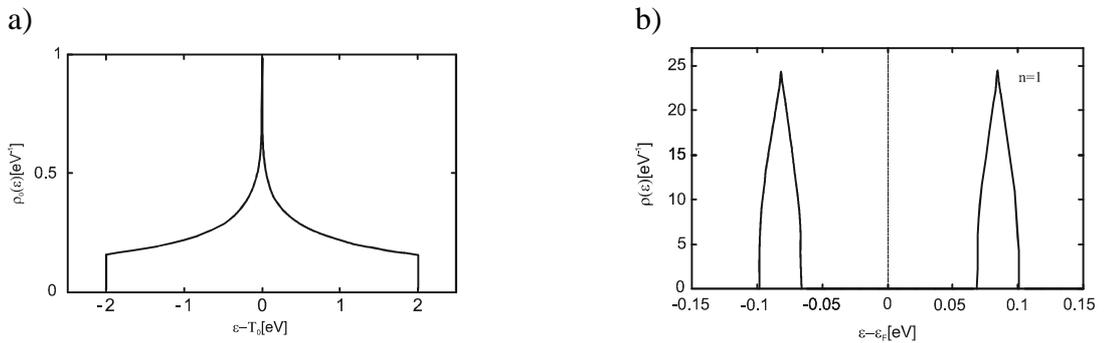

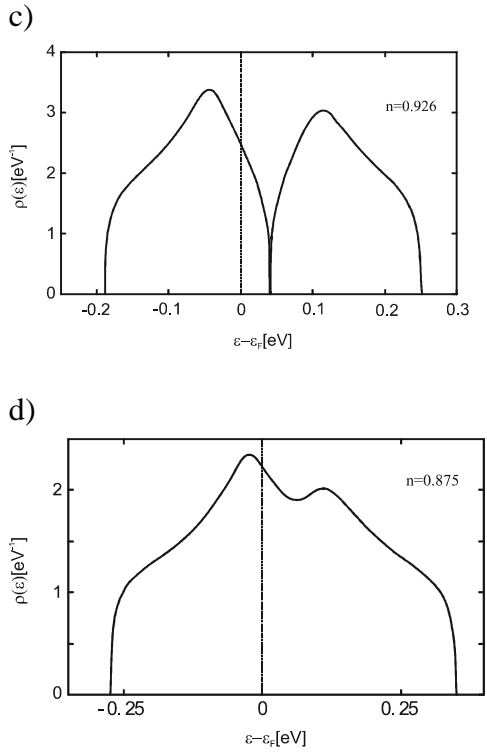

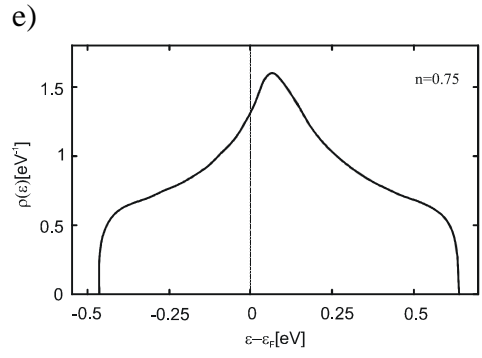

Fig. 8 The initial and modified by interactions densities of states versus energy relative to the Fermi level. Fig. (a) – initial two dimensional tight binding DOS without interactions ($T_0$ is the atomic energy). Figs (b)-(e): modified DOS for different values of carrier concentration, $n$. The values of interactions are: $F_{in} = 0.1$eV, $V \approx -0.18$eV, $U = 0.17$eV, $\Delta t = 0.99t$.

In Fig. 9 below we compare (as in the case of H-F approximation) the influence of $\Delta t$ interaction on the range of existence of the AF state.

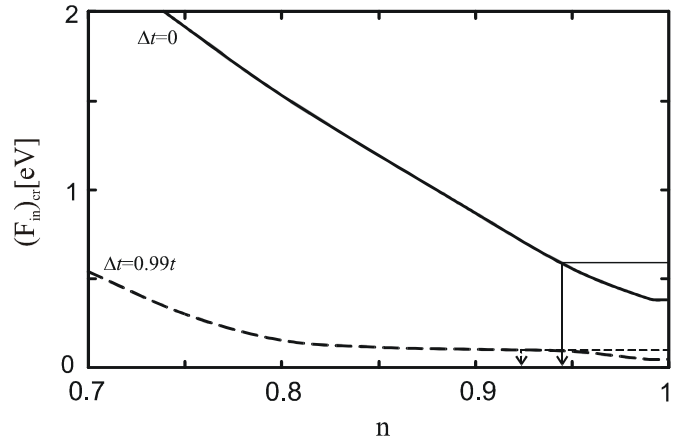

Fig. 9 The critical curves for the exchange on-site antiferromagnetic interaction $F_{in}$ in function of carrier concentration $n$ for different values of $\Delta t$ interaction. The Coulomb on-site repulsion ($U = 0.17 eV$) is treated in the CP approximation. The $F_{in}$ values fitted to the Néel's temperature at $n = 1$ are shown by the horizontal lines, which at the intersection with the critical curve determine the carrier concentration at which the AF state vanishes. The other parameters are: $V = -0.21$eV and $D_0 = 2$eV. The effective bandwidth is $D = D_0(1-\gamma n)$, $\gamma = \Delta t / t$.

As we can see, similarly to the H-F approximation, the $\Delta t$ interaction has negative influence on AF state by shifting it to the smaller concentrations in the case of initial half-bandwidth of the order of 2eV. The shift is relatively small but the influence of $\Delta t$ on SC state is significant (see Figs 10 and 12 below). On the other side the high value of the effective Coulomb repulsion given by $U/D(n)$ (due to a very small value of effective bandwidth $D(n)$ created by $\Delta t$ interaction at $n=1$) affects the SC state positively. The SC critical curve is pushed away from the half filling point (see Fig. 10), what brings it close to the experimental data for YBCO compound.

We can "adjust" the SC curve in two ways. One way is increasing the hopping interaction what causes the increase of the effective Coulomb repulsion ($U/D(n)$) and, in the result, shifts the SC phase towards smaller concentrations. Another way is by increasing the absolute value of charge-charge interaction $V$, what causes the growth of the whole SC curve.

In Fig. 10 below we present superconducting critical temperatures $T_{cr}^{SC}$ as a function of carrier concentration $n$ for different values of inter-site charge-charge interaction $V$.

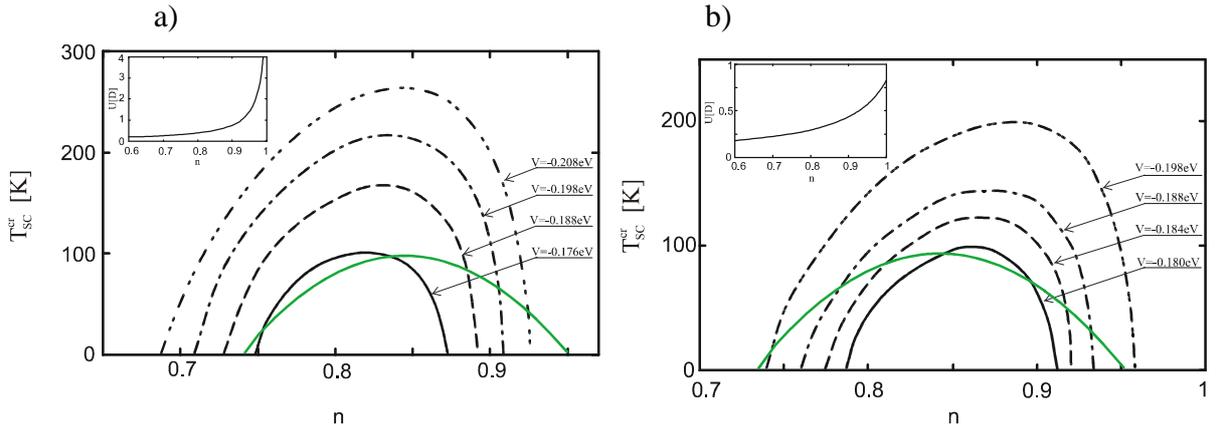

Fig. 10 Comparison of superconducting critical temperatures $T_{cr}^{SC}$ in function of carrier concentration $n$ for different values of inter-site charge-charge interaction $V$, with experimental data (green curves). The values of other interactions used in computations are: $F_{in}=0.1$eV, and $\Delta t = 0.99 t$; b) $F_{in}=0.15$eV, and $\Delta t = 0.9 t$. The half-bandwidth $D_0 = 2$eV and the Coulomb repulsion $U = 0.17$eV. The solid lines correspond to the red dashed lines in Fig. 10. Inset: the dependence of effective Coulomb repulsion $U[D(n)]$ on carrier concentration $n$, in units of the effective bandwidth; $D(n) = D_0(1-\gamma n)$, $\gamma = \Delta t / t$.

As can be seen, the size of the superconducting effect is very sensitive to the strength of negative charge-charge interaction. To justify the negative sign of this interaction we remind here that the effective single-band model came about by reducing the three-band extended Hubbard model representing $CuO_2$ planes in the high-$T_C$ cuprates. In this process the inter-site charge-charge interaction, $V$, was expressed as the difference between $V_z$ and $V_x$ (see Eq. (2)), where $V_z$ and $V_x$ are the interactions between the copper $3d_z$ and $3d_x$ orbitals and the oxygen $y$ direction orbital $p_y$ in the $CuO_2$ plane, respectively.

As was mentioned above, the hopping interaction affects the range of AF state mainly by changing the effective bandwidth. The complicated behavior of the density of states with respect to the changes of carrier concentrations and the hopping integral causes the complexity of the image presented in Fig. 11. This figure shows the influence of the hopping interaction on the Neel's temperature. The increase of $\Delta t$ (from 0 to $0.8t$) causes the shift of the AF curve to lower carrier concentrations. At $\Delta t = 0.8t$ the cavity starts to form in the middle of effective DOS. This cavity is caused by the relatively high value of ratio $U/D(n)$ (~0.8 at $n = 1$). When $\Delta t$ grows further, the cavity becomes bigger and at $\Delta t \sim 0.93t$ the DOS splits into two sub-bands at $n = 1$, and the AF curves "return" to the higher carrier concentrations. Another effect of increasing the $\Delta t$ interaction is the strong reduction of internal exchange field $F_{in}$, which is necessary to add for obtaining the antiferromagnetic alignment close to the half filled point.

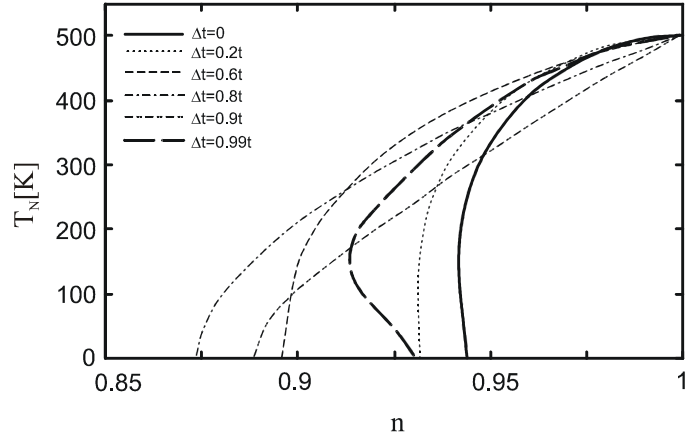

Fig. 11 The critical AF temperature in function of carrier concentration for different values of $\Delta t$. The half-bandwidth $D_0 = 2 \text{eV}$, $V = -0.178 \text{eV}$ and $U = 0.17 \text{eV}$. The values of $F_{in}$ were fitted to obtain correct maximum Neel's temperature, i.e.: $F_{in} = 0.1 \text{eV}$ for $\Delta t = 0.99t$, $F_{in} = 0.145 \text{eV}$ for $\Delta t = 0.9t$, $F_{in} = 0.207 \text{eV}$ for $\Delta t = 0.8t$, $F_{in} = 0.317 \text{eV}$ for $\Delta t = 0.6t$, $F_{in} = 0.507 \text{eV}$ for $\Delta t = 0.2t$, $F_{in} = 0.593 \text{eV}$ for $\Delta t = 0$.

In Fig. 12 we show the results for SC and AF in a full model of Eq. (1), i.e. with all interactions $U$, $V$, $F_{in}$ and $\Delta t$ being nonzero, and the on-site repulsion $U$ treated in the CP approximation. Our results (black and red curves) are compared with experiment (green curves). We can see that the agreement between theory and experiment is significantly improved as compared to Fig. 6 which was obtained with $\Delta t = 0$.

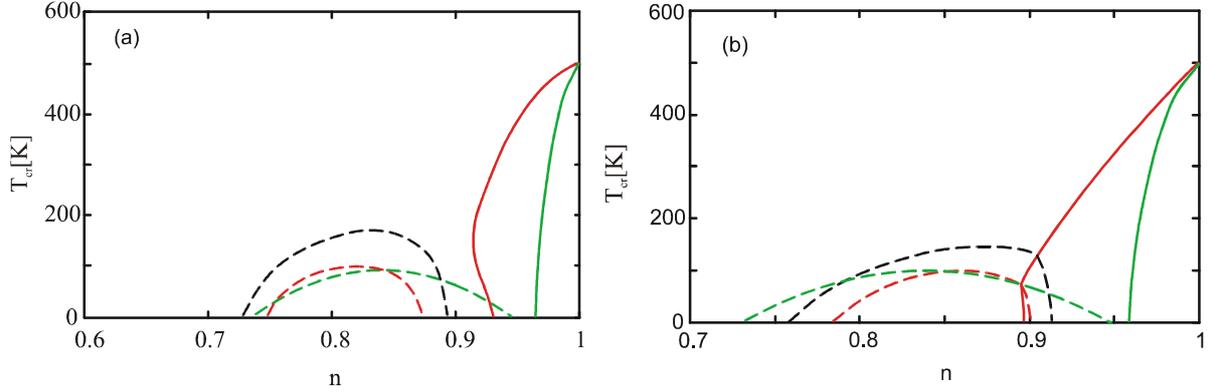

Fig. 12 Comparison of numerical results and experimental data (green curves) for the $YBa_2Cu_3O_7$ compound. The half-bandwidth $D_0 = 2\text{eV}$ and the Coulomb interaction is $U = 0.17\text{eV}$. The values of interactions used in the computations are: Fig. (a) $F_{in} = 0.1\text{eV}$, $V = -0.176\text{eV}$, and $\Delta t = 0.99t$; Fig. (b) $F_{in} = 0.15\text{eV}$, $V = -0.18\text{eV}$, and $\Delta t = 0.9t$. The black dashed line is for $V = -0.19\text{eV}$ (both diagrams). The critical AF curve (red line) is the same for all interactions $V$.

The hopping interaction, $\Delta t$, does not preserve the electron-hole symmetry. Therefore in Fig. 13 we show the critical temperatures $T_{SC}$ and $T_N$ for the hole-doped ($n<1$) and electron-doped ($n>1$) compounds at smaller values of the hopping interaction $\Delta t = 0.4t$.

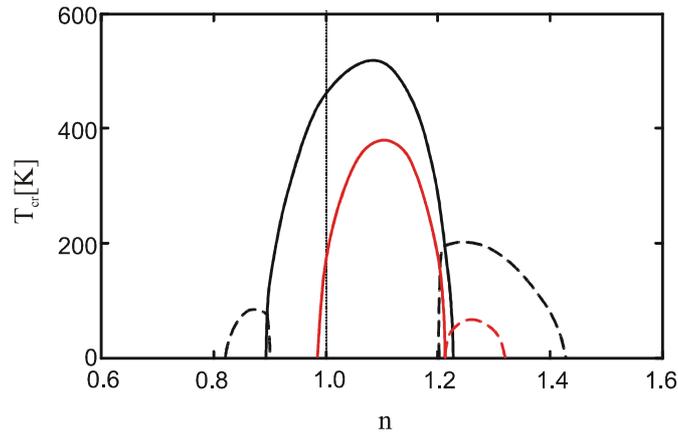

Fig. 13 The critical AF temperature (solid lines) and SC temperature (dashed lines) in function of carrier concentration. Black curves are for $V = -0.42\text{eV}$, and $F_{in} = 0.43\text{eV}$. Red curves are for $V = -0.4\text{eV}$, and $F_{in} = 0.41\text{eV}$. The other parameters the same for both curves are: $D_0 = 2\text{eV}$, $\Delta t = 0.4t$ and $U = 0.4\text{eV}$.

The figure shows that the stimulation of AF and SC by the hopping interaction is stronger for electron-doped ($n>1$) compounds. At smaller charge-charge interactions (the red curves) it is even possible for a given compound not to have a SC at the hole-doped concentrations ($n<1$), and to obtain a good fit to the experimental results at $n>1$ for the case of Nd$_{2-x}$Ce$_x$CuO$_{4-y}$ compound, see Fig. 1. The detailed numerical results with included coexistence between AF and SC ordering will be presented in the future.

**IV. Conclusions**

The results for the SC and AF critical temperatures after applying the CP approximation are close to the experimental data. Fast decrease of AF observed experimentally is obtained for relatively high initial bandwidth and the effective bandwidth increasing with carrier concentration departing from the half filled point (due to nonzero $\Delta t$ interaction). Mutual dumping of AF and SC states is strong, specially at relatively strong $U/D$, which is achieved at nonzero $\Delta t$.

In the Hartree-Fock approximation the Coulomb on-site repulsion $U$ does not influence SC directly. Indirectly, increase of $U$ increases Néel's temperature and the range of AF, which in turn is dumping slowly the SC state. Coexistence of SC and AF takes place in the broad range of concentrations.

In the CP approximation there is a relation between the magnitude of Coulomb repulsion $U$ and the SC critical temperature through the shape modification of initial DOS. The concentration dependence of the effective bandwidth, $D(n)$, due to the nonzero value of the $\Delta t$ interaction, enhances the effective Coulomb repulsion by increasing the $U/D(n)$ ratio at $n \to 1$. At this relatively strong Coulomb repulsion there is large shape modification of the initial DOS resulting in repulsion of the SC state away from the half filled point ($n=1$).

The increase of the initial bandwidth causes the range of AF state to shift towards larger concentrations closer to $n=1$. As mentioned just above the $U/D(n)$ ratio, increased by $\Delta t$ interaction, repels the SC to smaller concentrations. As a result at relatively high values of $\Delta t$ interaction there is no region of coexistence between SC and AF states in the CP approximation (see e.g. Fig. 10a). In this approximation, and when $\Delta t \neq 0$, both the $d$-wave SC critical curve and AF critical curve can be well adjusted to the experimental data for YBCO ceramics.

For electron-doped ($n>1$) compounds the hopping interaction stimulation of AF and SC is stronger than for the hole-doped ($n<1$) compounds. It is possible for a given compound not to have a SC at the hole-doped concentrations ($n<1$), and to obtain a good fit to the experimental results at $n>1$, see Fig. 1 for the experimental data for Nd$_{2-x}$Ce$_x$CuO$_{4-y}$ compound, and Fig. 13 for theoretical curves.

## Appendix A: Hartree-Fock Approximation for SC coexisting with AF

Treating the interaction $U$ in the H-F approximation we can write the self-energies appearing in Eq. (5) in the following form

$$\Sigma^{\sigma}_{\alpha(\beta)} = -F_{in} n^{\alpha(\beta)}_{\sigma} + U n^{\beta(\alpha)}_{\sigma}. \tag{A.1}$$

The self-energy described by above equation is energy independent. It depends on carrier concentration and magnetization only. Taking into account Eqs (9) and (10) we can calculate the non-magnetic and magnetic part of the self-energy. As a result we have

$$\Sigma_0 \equiv \frac{\Sigma^{\sigma}_{\alpha} + \Sigma^{-\sigma}_{\alpha}}{2} = (U - F_{in})\frac{n}{2}, \tag{A.2}$$

and

$$\Sigma_1 \equiv \frac{\Sigma^{\sigma}_{\alpha} - \Sigma^{-\sigma}_{\alpha}}{2} = -(U + F_{in})\frac{m}{2}. \tag{A.3}$$

After transforming Hamiltonian $H_0$ (Eq. (4)) and $H_{SC}$ (Eq. (6)) into the momentum space and solving the equations of motion for the Green's functions we obtain

$$\begin{bmatrix} \varepsilon - \varepsilon^{AF}_k + \mu - \Sigma_0 & \Delta_k & \Sigma_1 & 0 \\ \Delta^*_k & \varepsilon + \varepsilon^{AF}_k - \mu + \Sigma_0 & 0 & \Sigma_1 \\ \Sigma_1 & 0 & \varepsilon - \varepsilon^{AF}_{k+Q} + \mu - \Sigma_0 & \Delta_{k+Q} \\ 0 & \Sigma_1 & \Delta^*_{k+Q} & \varepsilon + \varepsilon^{AF}_{k+Q} - \mu + \Sigma_0 \end{bmatrix} \cdot \hat{G}(k,\varepsilon) = \hat{1}, \tag{A.4}$$

where the Green's functions are defined by the matrix elements

$$\hat{G}(k,\varepsilon) = \begin{pmatrix} \ll c_{k\uparrow};c^+_{k\uparrow} \gg & \ll c_{k\uparrow};c_{-k\downarrow} \gg & \ll c_{k\uparrow};c^+_{k+Q\uparrow} \gg & \ll c_{k\uparrow};c_{-k-Q\downarrow} \gg \\ \ll c^+_{-k\downarrow};c^+_{k\uparrow} \gg & \ll c^+_{-k\downarrow};c_{-k\downarrow} \gg & \ll c^+_{-k\downarrow};c^+_{k+Q\uparrow} \gg & \ll c^+_{-k\downarrow};c_{-k-Q\downarrow} \gg \\ \ll c_{k+Q\uparrow};c^+_{k\uparrow} \gg & \ll c_{k+Q\uparrow};c_{-k\downarrow} \gg & \ll c_{k+Q\uparrow};c^+_{k+Q\uparrow} \gg & \ll c_{k+Q\uparrow};c_{-k-Q\downarrow} \gg \\ \ll c^+_{-k-Q\downarrow};c^+_{k\uparrow} \gg & \ll c^+_{-k-Q\downarrow};c_{-k\downarrow} \gg & \ll c^+_{-k-Q\downarrow};c^+_{k+Q\uparrow} \gg & \ll c^+_{-k-Q\downarrow};c_{-k-Q\downarrow} \gg \end{pmatrix}, \tag{A.5}$$

and

$$\varepsilon^{AF}_k = \varepsilon_k b^{AF}, \qquad b^{AF} = 1 - \frac{\Delta t}{t} n + \frac{V}{t} I_{AF}. \tag{A.6}$$

The nesting vector $\mathbf{Q}$ has the property: $e^{-i\mathbf{Q}\cdot\mathbf{R}_{i\alpha}} = 1$, $e^{-i\mathbf{Q}\cdot\mathbf{R}_{i\beta}} = -1$ [in the case of two-dimensional square lattice we have $\mathbf{Q} = (\pi,\pi)$].

The Green's functions $\ll A;B \gg_\varepsilon$ calculated from the system of equations (A.4) allow us to find the averages

$$\langle BA \rangle = -\frac{1}{\pi} \int f(\varepsilon) \, \text{Im} \ll A; B \gg_\varepsilon d\varepsilon \,, \tag{A.7}$$

where $f(\varepsilon)$ is the Fermi distribution function.

These averages are necessary for calculating the carrier concentration, antiferromagnetic magnetization and the kinetic correlation parameter $I_{AF}$ which are given by the relations

$$n = \frac{1}{2N} \sum_{k,\sigma} \left( \langle c^+_{k\sigma} c_{k\sigma} \rangle + \langle c^+_{k+Q,\sigma} c_{k+Q,\sigma} \rangle \right), \tag{A.8}$$

$$m = \frac{1}{N} \sum_{k,\sigma} \sigma \langle c^+_{k\sigma} c_{k+Q,\sigma} \rangle, \tag{A.9}$$

$$I_{AF} = \frac{1}{N} \sum_{k,\sigma} \gamma_k \left( \langle c^+_{k\sigma} c_{k\sigma} \rangle - \langle c^+_{k+Q,\sigma} c_{k+Q,\sigma} \rangle \right), \tag{A.10}$$

where $\gamma_k \equiv \cos k_x + \cos k_y$ and $N$ is the number of lattice sites.

As a result we obtain the following equations for these quantities

$$n = 1 - \frac{1}{2N} \sum_k \left[ \frac{E_k^{AF} - \tilde{\mu}}{\tilde{E}_{k,2}} \tanh\left(\frac{\beta}{2} \tilde{E}_{k,2}\right) - \frac{E_k^{AF} + \tilde{\mu}}{\tilde{E}_{k,1}} \tanh\left(\frac{\beta}{2} \tilde{E}_{k,1}\right) \right], \tag{A.11}$$

$$m = \frac{\Sigma_1}{4N} \sum_k \left[ \left(1 + \frac{\tilde{\mu}}{E_k^{AF}}\right) \frac{\tanh\left(\frac{\beta}{2} \tilde{E}_{k,1}\right)}{\tilde{E}_{k,1}} + \left(1 - \frac{\tilde{\mu}}{E_k^{AF}}\right) \frac{\tanh\left(\frac{\beta}{2} \tilde{E}_{k,2}\right)}{\tilde{E}_{k,2}} \right], \tag{A.12}$$

$$I_{AF} = -\frac{1}{2N} \sum_k \gamma_k \varepsilon_k \left[ \left(1 + \frac{\tilde{\mu}}{E_k^{AF}}\right) \frac{\tanh\left(\frac{\beta}{2} \tilde{E}_{k,1}\right)}{\tilde{E}_{k,1}} + \left(1 - \frac{\tilde{\mu}}{E_k^{AF}}\right) \frac{\tanh\left(\frac{\beta}{2} \tilde{E}_{k,2}\right)}{\tilde{E}_{k,2}} \right], \tag{A.13}$$

where

$$\tilde{E}_{k,1(2)} = \sqrt{\left(\mp E_k^{AF} - \tilde{\mu}\right)^2 + \Delta_k^2} \,, \tag{A.14}$$

$$E_k^{AF} = \sqrt{\left(\varepsilon_k^{AF}\right)^2 + \Sigma_1^2} \,, \tag{A.15}$$

$$\tilde{\mu} = \mu - \Sigma_0 \,. \tag{A.16}$$

The equation for the *d*-wave superconducting state can be obtained by using a Fourier transform of the superconducting order parameter given by Eq. (11) into the momentum space. The result can be written as

$$\Delta_k = d\eta_k \,, \tag{A.17}$$

where the *d*-wave symmetry parameter $\eta_k = \cos k_x - \cos k_y$, $d = -2V\Delta_D$, and the parameter $2\Delta_D = \Delta_{\alpha,\alpha+x} - \Delta_{\alpha,\alpha+y}$ (index $\alpha + x(y)$ means the nearest neighbor of atom $\alpha$ in the $x(y)$ direction) has the following form

$$2\Delta_D = \frac{1}{2N}\sum_{k\sigma}\sigma\eta_k \langle c_{-k-\sigma}c_{k\sigma}\rangle$$
$$= \frac{1}{2N}\sum_{k\sigma}\sigma\eta_k \int f(\varepsilon)\left(-\frac{1}{\pi}\right)\text{Im} \ll c_{k\sigma}; c_{-k-\sigma} \gg d\varepsilon \qquad (A.18)$$

Next we use the moments method (see [17,20]), which in this case is reduced to comparing coefficients at terms with the second power of $\eta_k$ in Eq. (A.17). As a result we obtain the equation which allow us to obtain the *d*-wave superconducting critical temperature

$$1 = -VL_2, \qquad (A.19)$$

where the second-order moment is given by

$$L_2 = \frac{1}{4N}\sum_k \eta_k^2 \left[\frac{\tanh\left(\frac{\beta}{2}\tilde{E}_{k,1}\right)}{\tilde{E}_{k,1}} + \frac{\tanh\left(\frac{\beta}{2}\tilde{E}_{k,2}\right)}{\tilde{E}_{k,2}}\right]. \qquad (A.20)$$

The critical temperature is found from equation (A.19) in the limit of $\Delta_k \to 0$.

**Appendix B: Coherent Potential Approximation for SC coexisting with AF**

We treat the interaction $U$ in the coherent potential approximation. The interaction part of the Hamiltonian (5) takes on the following form

$$H_I = \sum_{i\sigma}\left(\tilde{V}_{i\alpha}^\sigma - \Sigma_\alpha^\sigma\right)\hat{n}_{i\sigma}^\alpha + \sum_{i\sigma}\left(\tilde{V}_{i\beta}^\sigma - \Sigma_\beta^\sigma\right)\hat{n}_{i\sigma}^\beta, \qquad (B.1)$$

with the stochastic potentials and the corresponding probabilities given by

$$\tilde{V}_{i\alpha(\beta)}^\sigma = \begin{cases}\tilde{V}_{1\alpha(\beta)}^\sigma = -F_{in}n_\sigma^{\alpha(\beta)} & P_{1\alpha(\beta)}^\sigma = 1 - n_{-\sigma}^{\alpha(\beta)} \\ \tilde{V}_{2\alpha(\beta)}^\sigma = U + \tilde{V}_{1\alpha(\beta)}^\sigma & P_{2\alpha(\beta)}^\sigma = n_{-\sigma}^{\alpha(\beta)}\end{cases}. \qquad (B.2)$$

They fulfill the following system of equations for the self-energies $\Sigma_\gamma^{\pm\sigma}(\varepsilon)$

$$\sum_{i=1}^{2} P_{i\gamma}^\sigma \frac{\tilde{V}_{i\gamma}^\sigma - \Sigma_\gamma^\sigma(\varepsilon)}{1 - \left[\tilde{V}_{i\gamma}^\sigma - \Sigma_\gamma^\sigma(\varepsilon)\right]F_\gamma^\sigma(\varepsilon)} = 0, \qquad \gamma = \alpha, \beta, \qquad (B.3)$$

with the Slater-Koster functions $F_{\alpha(\beta)}^{\sigma}(\varepsilon)$ equal to

$$F_{\alpha(\beta)}^{\sigma}(\varepsilon) = -\frac{1}{N}\sum_k \frac{\Delta_k^2\left(\varepsilon - \mu + \Sigma_{\alpha(\beta)}^{\sigma}\right) - \left(\varepsilon + \mu - \Sigma_{\beta(\alpha)}\right)\left[\left(\varepsilon - \mu + \Sigma_{\alpha}^{\sigma}\right)\left(\varepsilon - \mu + \Sigma_{\beta}^{\sigma}\right) - \left(\varepsilon_k^{AF}\right)^2\right]}{\left(\varepsilon^2 - E_{k,1}^2\right)\left(\varepsilon^2 - E_{k,2}^2\right)},$$

(B.4)

and

$$E_{k,1(2)} = \sqrt{\left(\mp E_k^{AF} - \mu + \Sigma_0\right)^2 + \Delta_k^2}\,, \quad E_k^{AF} = \sqrt{\left(\varepsilon_k^{AF}\right)^2 + \Sigma_1^2}\,, \quad \Sigma_{0(1)} \equiv \Sigma_{0(1)}(\varepsilon) = \frac{\Sigma_{\alpha}^{\sigma}(\varepsilon) \pm \Sigma_{\alpha}^{-\sigma}(\varepsilon)}{2}.$$

Solving this system of equations we obtain the following expressions for carrier concentration, antiferromagnetic magnetization and the kinetic correlation $I_{AF}$ parameter

$$n = 1 - \frac{1}{N}\sum_k \int f(\varepsilon)\frac{1}{\pi}\operatorname{Im} X_k(\varepsilon)d\varepsilon\,,$$  (B.5)

$$m = -\frac{1}{2N}\sum_k \int f(\varepsilon)\frac{1}{\pi}\operatorname{Im} Y_k(\varepsilon)d\varepsilon\,,$$  (B.6)

$$I_{AF} \equiv \left\langle c_{j\sigma}^+ c_{i\sigma}\right\rangle = -\frac{1}{N}\sum_k \gamma_k \int f(\varepsilon)\frac{1}{\pi}\operatorname{Im} Z_k(\varepsilon)d\varepsilon\,,$$  (B.7)

where the quantities $X_k(\varepsilon)$, $Y_k(\varepsilon)$, and $Z_k(\varepsilon)$ (added for clarity) are

$$X_k(\varepsilon) = -(\mu - \Sigma_0)\left[\frac{1}{\varepsilon^2 - E_{k,1}^2}\left(1 + \frac{E_k^{AF}}{\mu - \Sigma_0}\right) + \frac{1}{\varepsilon^2 - E_{k,2}^2}\left(1 - \frac{E_k^{AF}}{\mu - \Sigma_0}\right)\right],$$ (B.8)

$$Y_k(\varepsilon) = -\Sigma_1\left[\frac{1}{\varepsilon^2 - E_{k,1}^2}\left(1 + \frac{\mu - \Sigma_0}{E_k^{AF}}\right) + \frac{1}{\varepsilon^2 - E_{k,2}^2}\left(1 - \frac{\mu - \Sigma_0}{E_k^{AF}}\right)\right],$$ (B.9)

$$Z_k(\varepsilon) = \varepsilon_k\left[\frac{1}{\varepsilon^2 - E_{k,1}^2}\left(1 + \frac{\mu - \Sigma_0}{E_k^{AF}}\right) + \frac{1}{\varepsilon^2 - E_{k,2}^2}\left(1 - \frac{\mu - \Sigma_0}{E_k^{AF}}\right)\right],$$ (B.10)

$f(\varepsilon) = \{1 + \exp[\beta(\varepsilon - \mu)]\}^{-1}$ is the Fermi distribution function, $\beta = 1/(k_B T)$, and the symmetry parameter $\gamma_k = \cos k_x + \cos k_y$. The expressions given above are energy-dependent also through the self-energies ($\Sigma_{0(1)} \equiv \Sigma_{0(1)}(\varepsilon)$).

Using the expression for the superconducting order parameter, $\Delta_k$ (see Eq. (A.17)), and proceeding in analogous way to the case of H-F approximation, we arrive at equation for the $d$-wave superconducting state in the following form

$$1 = -VL_2,$$  (B.11)

where the moment of the second order, $L_2$, is given by

$$L_2 = \frac{1}{2N}\sum_k \eta_k^2 \iint \left[S_1(k,\varepsilon_1)S_1(k,\varepsilon_2) + S_2(k,\varepsilon_1)S_2(k,\varepsilon_2)\right]\frac{\tanh(\beta\varepsilon_1/2) + \tanh(\beta\varepsilon_2/2)}{2(\varepsilon_1+\varepsilon_2)}d\varepsilon_1 d\varepsilon_2,$$

(B.12)

with the spectral functions $S_{1(2)}(k,\varepsilon) = -\frac{1}{\pi}\operatorname{Im}\frac{1}{\varepsilon - E_{k,1(2)}(\varepsilon)}$, and the $d$-wave symmetry parameter $\eta_k = \cos k_x - \cos k_y$.

To obtain SC critical temperature we solve Eq. (B.11) in the limit of $\Delta_k \to 0$ together with Eqs (B.5)-(B.7), and with the self energies calculated from the Eq. (B.3).